\documentclass[12pt]{article}
\usepackage{psfig}
\begin{document}
\begin{center}

{\large \bf Detailed Study of the Ursa Major Supercluster
	    of Galaxies Using the 2MASS and SDSS Catalogs}

\vspace{0.8cm}
{\bf Flera Kopylova$^*$ and Alexander Kopylov}

{\it Special Astrophysical Observatory, Russian Academy of Sciences,
     Nizhnii Arkhyz, Karachai-Cherkessian Republic, 369167 Russia}

\begin{abstract}
We study the infrared ($K_{s}$ band) properties of clusters
of galaxies in the Ursa Major supercluster using data from 2MASS (Two-Micron
All-Sky Survey) and SDSS (Sloan Digital Sky Survey). We identified three
large filaments with mean redshifts of $z$ = 0.051, 0.060, and 0.071. All
clusters of the supercluster are located in these filaments. We determined
the total $K_{s}$-band luminosities and masses for 11 clusters of galaxies
within comparable physical regions (within a radius $R_{200}$ close to the
virial radius) using a homogeneous method. We constructed a combined
luminosity function for the supercluster in this  region, which can be
described by the Schechter function with the following parameters:
$M^{*}_{K}$ = $-24^{m}.50$ and $\alpha = -0.98$. The infrared luminosities
of the clusters of galaxies correlate with their masses; the $M/L_{K}$
ratios of the systems increase with their masses (luminosities), with most
of the Ursa Major clusters of galaxies (particularly the richer ones) closely
following the relations derived previously for a large sample of clusters
and groups of galaxies. The total mass-to-infrared-luminosity ratio is
52 $M_{\odot}/L_{\odot}$ for six Abell clusters and 49 $M_{\odot}/L_{\odot}$
for all of the clusters, except Anon2.

\vspace{0.8cm}
Key words: {\it galaxies, groups and clusters of galaxies.}
\end{abstract}

\vspace{1.5cm}
\renewcommand{\thefootnote}{$*$}
\footnotetext{E-mail: flera@sao.ru}
\end{center}

\newpage
\section{Introduction}
\hspace*{\parindent}
The Ursa Major (UMa) supercluster ($11^h30^m$ $+55^0$) is one of
the closest $(z\simeq0.06)$, compact (in projection onto the celestial sphere)
superclusters. The apparent overdensity of the system, if defined with
respect of the number of Abell clusters in the surrounding region of about
150 Mpc in size, is 30. Being fairly isolated (there are no large Abell
clusters nearby), the supercluster shows an example of how the systems of
galaxy clusters evolve in space and time in the absence of external effects.
Our studies (Kopylov and Kopylova 2001; Kopylova and Kopylov 2001) revealed
that the UMa system is not a supermassive supercluster like the Corona
Borealis system and the Shapley Concentration, and, in general, obeys
the Hubble radial velocity-distance relation.

In this paper, we used
data from two large surveys, 2MASS and SDSS. The main advantages of
infrared (IR) photometric studies is their relative insensitivity to dust
and to the last starburst; therefore, the stellar masses of the galaxies
are traced better. Recently, using different approaches and methods for
different samples of galaxy clusters and groups, Lin et al. (2003, 2004)
(93 systems), Rines et al. (2004) (9 systems), Ramella et al. (2004)
(55 systems), Karachentsev and Kut'kin (2005) (groups of galaxies) have
found an increase in $M/L_{K}$ for the clusters and groups with mass
(luminosity) of the system in the $K_{s}$ band. In this paper, we are going
to test all these conclusions using the UMa supercluster of galaxies as
an example --- not a random set of galaxy clusters, but a system, though
not a virialized one.

The paper is structured as follows. First, we
describe the technique for determining the virial mass of a galaxy
cluster using the galaxy velocity dispersion. Subsequently, we consider
the features of the large-scale structure in the region of the supercluster
using galaxies from the 2MASS XSC catalog and their $z$ measurements from
SDSS. Next, we describe the technique for determining the total IR
luminosities of clusters using 2MASS data; we present the relation
between the derived virial masses and IR luminosities and the relation
between the mass-to-luminosity ratio and the mass for the UMa clusters
of galaxies. In conclusion, we summarize our main results. The following
cosmological parameters are used in this paper: $\Omega_m = 0.3$,
$\Omega_{\Lambda}=0.7$, $H_0 = 70~km~s^{-1}~Mpc^{-1}$.

\section{Spectroscopic data}

The third data release of the SDSS catalog (DR3; Abazajian et al. 2005)
allowed us to compile a sample of galaxies in the supercluster region
($7^\circ.9\times3^\circ.2$) with 2MASS measurements. Previously (Kopylov
and Kopylova 2001; Kopylova and Kopylov 2001), we measured the radial
velocities only for $10-15$ galaxies in each cluster. We found 2480 2MASS
XSC galaxies in the field of the UMa supercluster: 109 of them are foreground
galaxies, 1011 galaxies belong to the supercluster ($0.045 < z < 0.075$),
939 are background galaxies, 421 galaxies have no measured $z$.

Before selecting the galaxies that belong to each of the clusters, we
determined the cluster centers. In general, these are the brightest galaxies
of the clusters (or one of the brightest galaxies, if there were a few of
them), except two cases: we selected the centers of the clusters A1270 and
A1436, respectively, between two concentrations of galaxies and between two
bright galaxies. X-ray radiation was detected rather confidently from
the clusters A1291, A1377, A1436, and Anon1 (Bohringer et al., 2000);
the clusters A1270, A1318, and A1383 were measured less reliably
(Ledlow et al., 2003). In all cases, the discrepancies between the centers
we selected and those determined from the X-ray radiation do not exceed
$4^{\prime}$.

For each cluster, we selected galaxies from SDSS and 2MASS XSC within
$30^{\prime}$ of the derived centers (for the supercluster's mean distance,
$1^{\prime}$ corresponds to 70 kpc). It turned out that, on average, the
number of galaxies in SDSS is larger by one third. The $K_{s}$ data for them
were taken from the 2MASS Point Source Catalog (PSC). We determined the mean
radial velocity of the cluster, $cz$, and its dispersion, $\sigma$,
iteratively: first, we used all of the galaxies with measured radial
velocities, except those deviating by more than $2.5\sigma$. Subsequently,
having assumed that all of the clusters are in virial equilibrium and their
masses  crease linearly with radius, we calculated in
$R_{200} = \sqrt{3}\sigma (1+z)^{-3/2}/(10H_{0})$ Mpc for the derived
velocity dispersion (Carlberg et al., 1997) and redetermined the cluster's
mean radial velocity, $cz$, and its dispersion, $\sigma_{200}$, within
$R_{200}$. $R_{200}$ is the cluster radius within which the density is a
factor of 200 higher than the critical density; it is approximately equal
to the virial radius. The virial mass within $R_{200}$ is
$M_{{vir},{200}} = 3G^{-1}R_{200} \sigma_{200}^2$ .

All of the data obtained for the clusters within R200 are presented in the
table. Column 1 gives the cluster names (Kopylov and Kopylova 2001);
columns 2 and 3 list the equatorial coordinates for the epoch J2000.0;
column 4 gives the cluster's heliocentric redshift (SDSS); column 5
contains the velocity dispersion within $R_{200}$, with the cosmological
correction $(1+z)^{-1}$ ; columns 6, 7, 8, and 9 present the number of
galaxies in the cluster with IR data and measured $z$, the radius of
the region under study, mass, total luminosity within $R_{200}$ (determined
below); and column 10 gives the mass-to-luminosity ratio for the cluster.

Figure 1 shows the integral distributions of galaxies in projected
distance from the cluster center normalized to $R_{200}$. The Abell clusters,
the richer ones in the UMa supercluster (Figs. 1a and 1b), have similar
distributions with galaxies within $R_{200}$, especially in the range of
distance $R/R_{200} = 0.35-0.6$ from their centers. The remaining clusters
(Figs. 1b and 1c) are poor, the agreement between them is poorer, and
the cluster Anon2 deviates significantly from the mean distribution.
The density of galaxies in this cluster is lower than that needed within
$R_{200}$ for virial equilibrium to occur.

\subsection{
The Large-Scale Structure in the Region of the Supercluster}

The measured radial velocities of galaxies (more than 1000 within the
supercluster) make it possible to study the large-scale structure in
the system. The distribution of galaxies is known to have a filamentary
structure: in general, the clusters are connected by filaments, and
the richest clusters are usually located at the points of intersection
between the filaments. Three large filaments that contain almost all of
the clusters are identified in the UMa supercluster. The galaxies
belonging to each of the filaments are indicated by different symbols
in Fig. 2. The large circles mark the virialized regions of the galaxy
clusters ($R_{200}$).

Data on the distribution of galaxies near clusters
make it possible to reveal cases where the field galaxies projected onto
the cluster can distort significantly the velocity dispersion. Since
the filaments in the UMa supercluster are located roughly in the plane
of the sky, do not overlap, and are fairly far apart along the line
of sight, the velocity dispersions and, hence, virial masses of the
clusters are determined rather reliably for all of the clusters,
except A1291.

The filament in the region of the cluster A1291, classified
by Abell as a cluster of richness class 1, is oriented roughly along
the line of sight. Several peaks in the radial-velocity distribution
from 14000 to 18500$~km~s^{-1}$ are present here (in this region, the data
completeness in $z$ in SDSS DR3 is lower than the mean). We determined
A1291 as the cluster related to the strongest peak (the velocity range
$14000-16000~km~s^{-1}$) that includes the cluster's brightest galaxy,
A1291$-$74 (Kopylov and Kopylova 2001). This is an unusual, strongly
deformed elliptical galaxy, a radio
source with a head-tail structure (Kopylov et al. 2005). The cluster Anon2
is in the immediate vicinity ($0^\circ.5$) of Anon1, has a low velocity
dispersion, and, contains mostly bright galaxies. Anon2 is most likely a
cluster that has not yet been formed completely, but is still at an early
virialization stage.

\section{Photometric data}

2MASS presents complete homogeneous all-sky photometry in three infrared
bands ($J$, $H$, $K_{s}$). $K_{s}$ is a modified K band;
below, we will denote it by omitting the subscript. We used the
photometry presented in the final version of the extended source catalog
(XSC; Jarrett et al. 2000) and in the point source catalog (PSC). From the
magnitudes of galaxies measured in XSC, we took, as recommended by
Jarrett et al. (2000), the isophotal magnitudes corresponding to the surface
brightness $\mu_{K} = 20~mag~arcsec^{-2}$ (we determined the total magnitudes
by subtracting $0^{m}.2$ from the isophotal magnitudes, as prescribed by
Kochanek et al. (2001)). The mean error of the isophotal magnitudes in our
sample is $0^{m}.1$.

The errors in the K magnitudes of galaxies in the PSC are rather large;
for example, they were estimated by Bell et al. (2003) to be $\sim0^{m}.5$.
In this paper, we found the difference between the $K_{20}$ (XSC) and
$K_{mstdap}$ (PSC) magnitudes for 80 galaxies (brighter than $13^{m}$ and
fainter than $13^{m}$); it turned out to be $-0^{m}.48\pm0^{m}.21$ and
$-0^{m}.13\pm0^{m}.17$, respectively. In addition, galaxies without measured
$z$ were found in the field of each cluster; on average, there are $1-4$ of
them, with 11 in A1436, 12 in A1377, 26 in A1291. Their cluster membership
was determined from the ($g-r, r$) color-magnitude diagram. Their contribution
to the luminosity was derived as being proportional to the fraction of the
cluster members with measured $z$ in the total number of 2MASS galaxies
with known $z$ within $R_{200}$.

\subsection{
The K -Band Luminosity of the Supercluster}

For the UMa supercluster, the photometric limit corresponding to the XSC
limit ($13^{m}.5$) is about $M^{*}_{K}+1$, $M^{*}_{K}$ where  is the
characteristic value of the field luminosity function (LF) equal to
$-24^{m}.16$ (Kochanek et al. 2001). To construct the LF, we converted the
observed galaxy magnitudes to their absolute magnitudes using the formula:
\begin{equation}
M_K = K-25-5 \log_{10}(D_{l}/1 Mpc)-A_K-K(z),
\end{equation}
where $D_{l}$ is the luminosity distance, $A_{K}$ is the extinction in the
Galaxy, and $K(z) = -6\log(1+z)$ (Kochanek et al. 2001) is the K-correction.
The galaxies were counted in $0^{m}.5$ and $0^{m}.25$ bins. To find the
parameters ($M^{*}_{K}, \alpha$) at $R_{200}$, we used the
Schechter (1976) function to fit the observed LF. We normalized the LF
to the total observed number of galaxies. Figure 3 is the combined LF for
the UMa supercluster. The curve indicates the derived Schechter function
with its parameters: $M^{*}_{K} = -24^{m}.50$, $\alpha = -0.98$. To
determine the total luminosity of the cluster galaxies within $R_{200}$,
we added the luminosities of all the individually measured galaxies by
taking $K_{\odot} = 3^{m}.32$ for the solar luminosity (Bell et al. 2003)
and then used the derived parameters of the Schechter function for
extrapolation to faint galaxies. We added the luminosities $L_{K,i}$ of
the observed members up to the catalog's limit ($13^{m}.5$) and integrated
the luminosity function in the region of faint magnitudes according to the
expression
\begin{equation}
L_{K}=\sum_{i=1}^{N_{obs}}L_{{K},i}+\phi^*L^*_{K}\int^{L_{lim}}
_{0}t^{\alpha+1} e^{-t} dt,
\end{equation}
where $t = L/L^{*}$.\\ Bright galaxies give a major contribution to the
cluster's luminosity. The calculated luminosities of the clusters correspond
to a cylinder of radius $R_{200}$. The correction to the spherical volume
reduces the cluster luminosities, on average, by 20\% (Ramella et al. 2004),
but this correction can be smaller, depending on the distribution of galaxies
in the vicinity of the cluster. In most cases the correction for the UMa
supercluster will be at a minimum, because the filaments are oriented in
the plane of the sky (Fig. 2).

\subsection{
Relations Between the IR Luminosity and Dynamical Parameters of the
Supercluster}

Figure 4 shows the correlation between the masses and luminosities of
the UMa clusters in the infrared (the K band). The filled circles
highlight the Abell clusters. The line corresponds to
$L_{200}/10^{12}L_{\odot} = 2.76 (M_{200}/10^{14}M_{\odot})^{0.72\pm0.04}$,
derived by Lin et al. (2004) for a sample of 93 clusters and groups of
galaxies with X-ray temperatures from 0.8 to $\sim12$ keV and masses from
$3\times10^{13}$ to $1.7\times10^{15}$ $M_{\odot}$. We found a similar
correlation between the
luminosity ($L_{K}$ ) and radial-velocity dispersion ($\sigma$):
$L_K \propto \sigma^{{1.90}\pm0.19}$. This relation for virialized structures
is estimates to be $L_B \sim\sigma^{2}$ (Girardi et al. 2000), in agreement
with our result. Figure 5 shows the dependence of the mass-toluminosity
ratio on the cluster mass. The dashed line is
$M_{200}/L_{200} \propto M^{0.26\pm0.04}_{200}$ (Lin et al. 2004). There is
also a similar dependence on the luminosity. The UMa supercluster consists
of 11 clusters with different properties, from very poor, in fact, groups
to rather rich; it can be noticed (Figs. 4 and 5) that the rich Abell
clusters (A1270, A1377, A1383, and A1436) deviate from the relation
by Lin et al. (2004) only slightly, whereas some of the poorer clusters
deviate significantly. If it is considered that we studied all clusters
in the same way, the masses and luminosities were estimated by assuming
virialization within $R_{200}$, then the deviations of some of the clusters
(open circles) in the diagram may stem from the fact that they have not
yet been virialized within this radius. The ratio of the total mass to
the total IR luminosity within $R_{200}$ for the Abell clusters (filled
circles) is 52 $M_{\odot}/L_{\odot}$ (49 for all of the clusters, except
Anon2). These values for the clusters forming the UMa supercluster are
close to those found (a) for the Coma cluster (A1656) within $R_{200}$:
$50\pm7$ $M_{\odot}/L_{\odot}$ (Rines et al. 2001); (b) for 27 clusters
of galaxies within $R_{500}$: $47\pm3$ $M_{\odot}/L_{\odot}$
(Lin et al. 2003); (c) for 8 CAIRNS clusters within $R_{200}$:
$49\pm5$ $M_{\odot}/L_{\odot}$ (Rines et al. 2004). The mass-to-infrared-
luminosity ratios derived from different ensembles of galaxy clusters is
always smaller than the universal ratio of $90\pm19$ $M_{\odot}/L_{\odot}$,
as determined by Lin et al. (2004) for $\Omega_{m}$ (WMAP) and the mean
luminosity density in the Universe measured by Bell et al. (2003).

\section{Conclusions}

As the most massive gravitationally bound and virialized (in the central
regions) objects in the Universe, the clusters of galaxies are the main
systems for studying the distribution of total dynamical mass and luminous
matter. The infrared radiation from galaxies is most suitable for these
purposes. For the UMa supercluster, which consists of 11 clusters with
masses from $0.3\times10^{14}$ to $6\times10^{14}$ $M_{\odot}$ and
luminosities from $1.5\times10^{12}$ to $9\times10^{12}$ $L_{\odot}$, we
determined the cluster masses within $R_{200}$ close to $R_{vir}$ and total
luminosities using the combined luminosity function of the clusters. Our
analysis of the relations between the mass and IR luminosity and between
the mass-toluminosity ratio and mass (Figs. 4, 5) and comparison of them
to the relations derived by Lin et al. (2003, 2004) for a large sample of
galaxy clusters shows that our data for the UMa cluster confirm the results
obtained by Lin et al. (2003, 2004) with a rather small scatter. Some,
mostly poorer clusters of galaxies exhibit appreciable deviations. This may
be because the system deviates from virial equilibrium (see, e.g., Ramella
et al. 2004), since the masses of the system is determined under this
assumption.

Our main conclusions are as follows.
(1) The five Abell clusters (except A1291) in the UMa supercluster
demonstrate similar integral distributions of their galaxies in projected
distance from the cluster center within the region under study, $R_{200}$.
The remaining clusters show various deviations. The strongest deviation
is found for the cluster Anon2, most probably because of its considerable
deviation from the dynamic equilibrium.
(2) The infrared ($K_{s}$-band) luminosities of the galaxy clusters
correlate with their masses; the $M/L_{K}$ ratio also increases with both
mass and luminosity, in a good agreement with the relations derived by
Lin et al. (2003, 2004) for a large sample of clusters and groups of galaxies.
(3) The total ratio of the mass of the Abell galaxy clusters to their IR
luminosity is $M_{\odot}/L_{\odot}$. For all the clusters, except Anon2,
this ratio is 49 $M_{\odot}/L_{\odot}$.

{\bf Acknowledgments}

This work was supported in part by the "Astronomy" Federal
Science and Technology Program.

\bigskip
{\bf References}\\
K. Abazajian, R. Abazajian, I. K. Adelman-McCarthy, et al.,
  Astron. J. {\bf129}, 1755 (2005).\\
E. F. Bell, D. H. McIntosh, N. Katz, and M. D. Weinberg,
  Astrophys. J., Suppl. Ser. {\bf149}, 289 (2003).\\
H. Bohringer, W. Voges, J. P. Huchra, et al.,
  Astrophys. J., Suppl. Ser. {\bf129}, 435 (2000).\\
R. G. Carlberg, H. K. C. Yee, E. Ellingson, et al.,
  Astrophys. J. {\bf485}, L13 (1997).\\
M. Girardi, S. Borgani, G. Guiricin, et al.,
  Astrophys. J. {\bf530}, 62 (2000).\\
T. H. Jarrett, T. Chester, R. Cutri, et al.,
  Astron. J. {\bf119}, 2498 (2000).\\
I. D. Karachentsev and A. M. Kut'kin,
  Astron. Lett. {\bf31}, 299 (2005).\\
C. S. Kochanek, M. A. Pahre, E. E. Falco, et al.,
  Astrophys. J. {\bf560}, 566 (2001).\\
A. I. Kopylov and F. G. Kopylova,
  Astron. Lett. {\bf27}, 140 (2001).\\
A. I. Kopylov and F. G. Kopylova,
  Astron. Lett. {\bf27}, 345 (2001).\\
A. I. Kopylov, F. G. Kopylova, and A. G. Gubanov,
  Astron. Astropys., submitted (2005).\\
M. J. Ledlow, W. Voges, F. N. Owen, and J. O. Burns,
  Astron. J. {\bf126}, 2740 (2003).\\
Y.-T. Lin, J. J. Mohr, and S. A. Stanford,
  Astrophys. J. {\bf591}, 749 (2003).\\
Y.-T. Lin, J. J. Mohr, and S. A. Stanford,
  Astrophys. J. {\bf610}, 745 (2004).\\
M. Ramella, W. Boschin, M. Geller, et al.,
  Astron. J. {\bf128}, 2022 (2004).\\
K. Rines, M. J. Geller, M. J. Kurtz, et al.,
  Astrophys. J. {\bf561}, L41 (2001).\\
K. Rines, M. J. Geller, A. Diaferio, et al.,
  Astron. J. {\bf128}, 1078 (2004).\\
P. Schechter,
  Astrophys. J. {\bf203}, 297 (1976).

\bigskip
\bigskip
{\it Translated by N. Samus'}

\newpage
\vspace{1.0cm}
\vbox{
\begin{center}

 Table. Data for clusters in the Ursa Major supercluster

\vspace{0.5cm}
\begin{tabular}{lccccccccc}
\hline \hline
Cluster& RA& DEC& z& $\sigma $& $N$& $R_{200}$& $M_{200}$& $L_{200}$ &$M/L_K$\\
 & J2000& J2000&  & $km~s^{-1}$&  & Mpc& $10^{14} M_{\odot}$& $10^{12} L_{\odot}$& $M_{\odot}/L_{\odot}$\\
 \hline
A1270& 11 29 42.0& +54 05 56& 0.06890& $553\pm76$& 53& 1.24& $2.65\pm0.73$& $5.57\pm0.16$& $48\pm13$\\
A1291& 11 32 23.2& +55 58 03& 0.05092&$424\pm103$& 17& 0.97& $1.22\pm0.60$& $2.37\pm0.10$& $52\pm25$\\
A1318& 11 36 03.5& +55 04 31& 0.05647& $411\pm67$& 37& 0.94& $1.11\pm0.36$& $3.41\pm0.17$& $33\pm11$\\
A1377& 11 47 21.3& +55 43 49& 0.05170& $613\pm71$& 74& 1.40& $3.68\pm0.86$& $6.61\pm0.12$& $56\pm13$\\
A1383& 11 48 05.8& +54 38 47& 0.05979& $527\pm71$& 55& 1.20& $2.33\pm0.62$& $4.66\pm0.13$& $50\pm13$\\
A1436& 12 00 08.8& +56 10 52& 0.06517& $701\pm83$& 71& 1.60& $5.50\pm1.30$& $9.34\pm0.14$& $59\pm16$\\
Anon1& 11 15 23.8& +54 26 39& 0.06944& $561\pm88$& 41& 1.25& $2.75\pm0.86$& $4.41\pm0.14$& $62\pm20$\\
Anon2& 11 19 46.0& +54 28 02& 0.07056& $253\pm70$& 13& 0.60& $0.27\pm0.15$& $2.86\pm0.16$&   $9\pm5$\\
Anon3& 11 29 32.3& +55 25 20& 0.06806& $362\pm70$& 23& 0.81& $0.74\pm0.29$& $2.40\pm0.16$& $31\pm12$\\
Anon4& 11 39 08.5& +55 39 52& 0.06118& $391\pm80$& 24& 0.89& $0.98\pm0.39$& $3.97\pm0.14$& $25\pm10$\\
Sh166& 12 03 11.9& +54 50 50& 0.04996& $318\pm76$& 18& 0.73& $0.52\pm0.24$& $1.45\pm0.09$& $36\pm17$\\
 \hline
\end{tabular}
\end{center}
}

\newpage
\begin{figure}[*]
\centerline{\psfig{figure=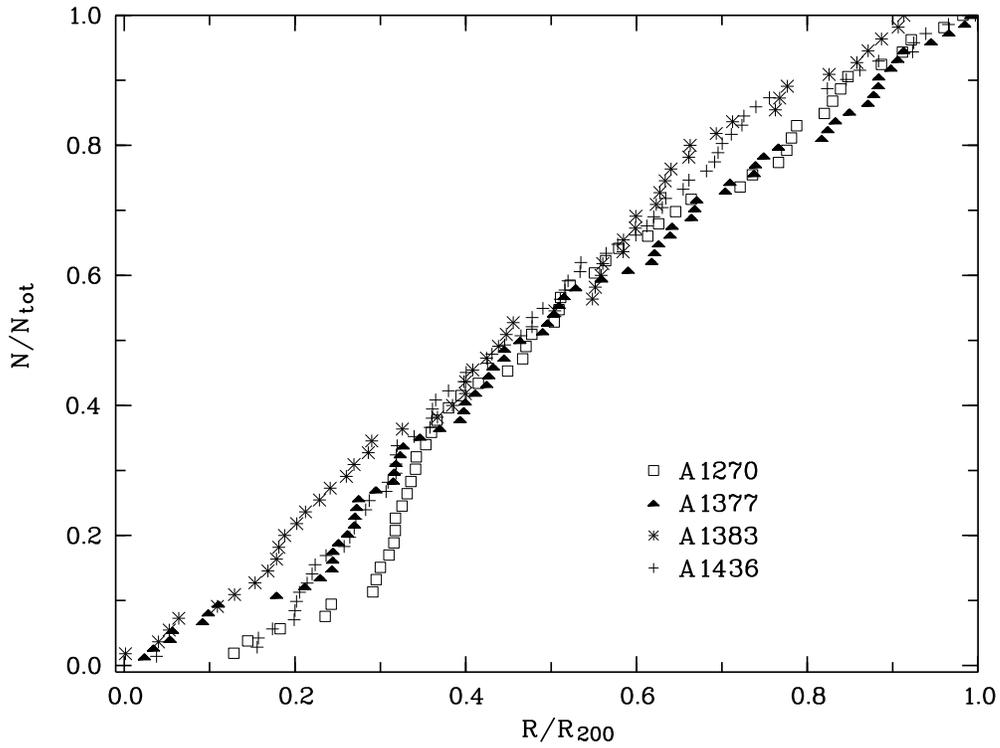,width=14cm,angle=-90,%
bbllx=69pt,bblly=133pt,bburx=520pt,bbury=730pt,clip=}}
\caption[]{Integral distributions of galaxies in projected
distance from the cluster center normalized to the total number
of galaxies within $R_{200}$. (a) A1270, A1377, A1383, A1436;
(b) A1318, Anon1, Anon4, Sh166; and (c) Anon2, Anon3, A1291.
}
\end{figure}
\begin{figure}[*]
\setcounter{figure}{0}
\centerline{\psfig{figure=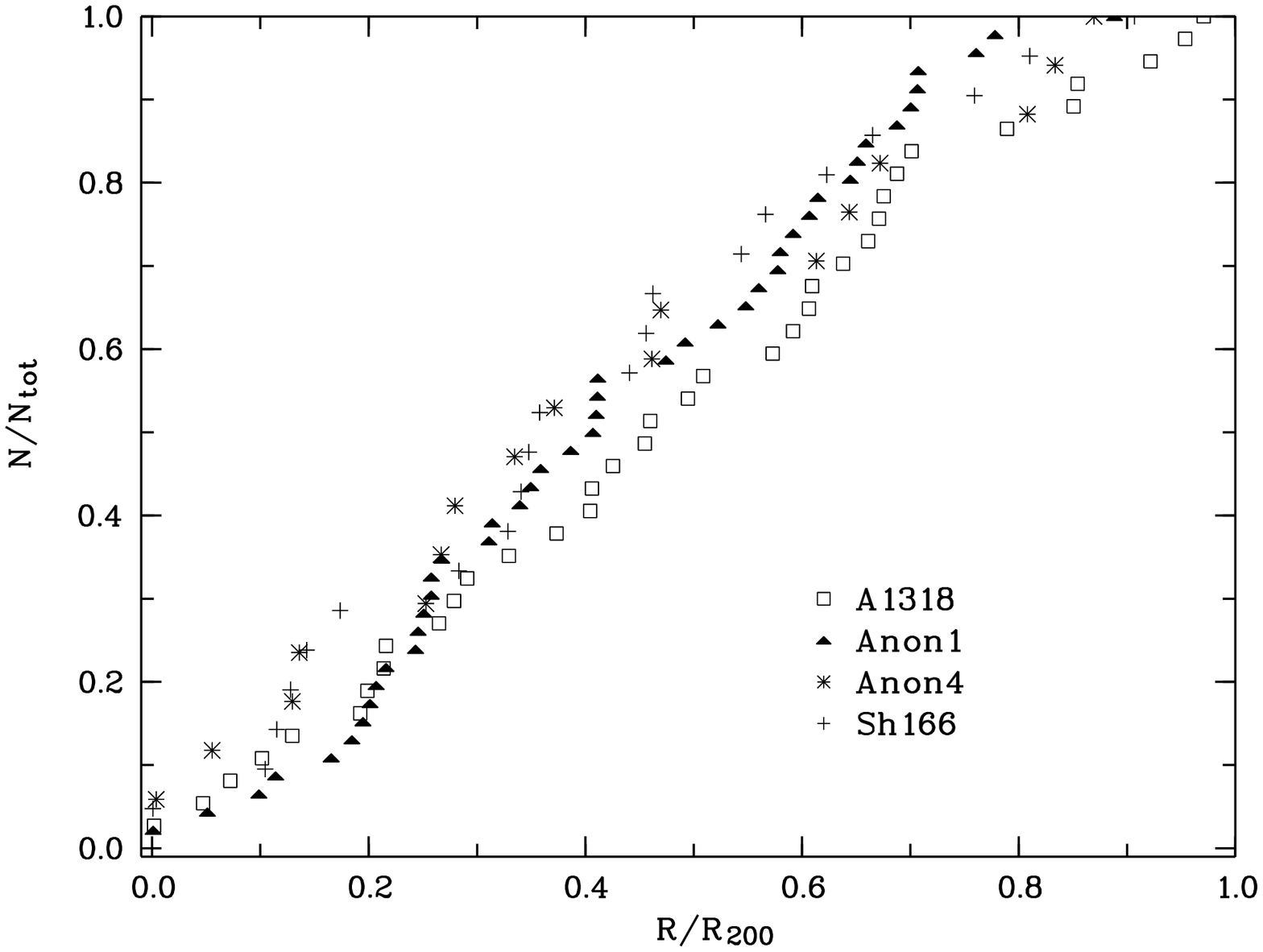,width=14cm,angle=-90,%
bbllx=69pt,bblly=133pt,bburx=520pt,bbury=730pt,clip=}}
\caption[]{(b)
}
\setcounter{figure}{0}
\centerline{\psfig{figure=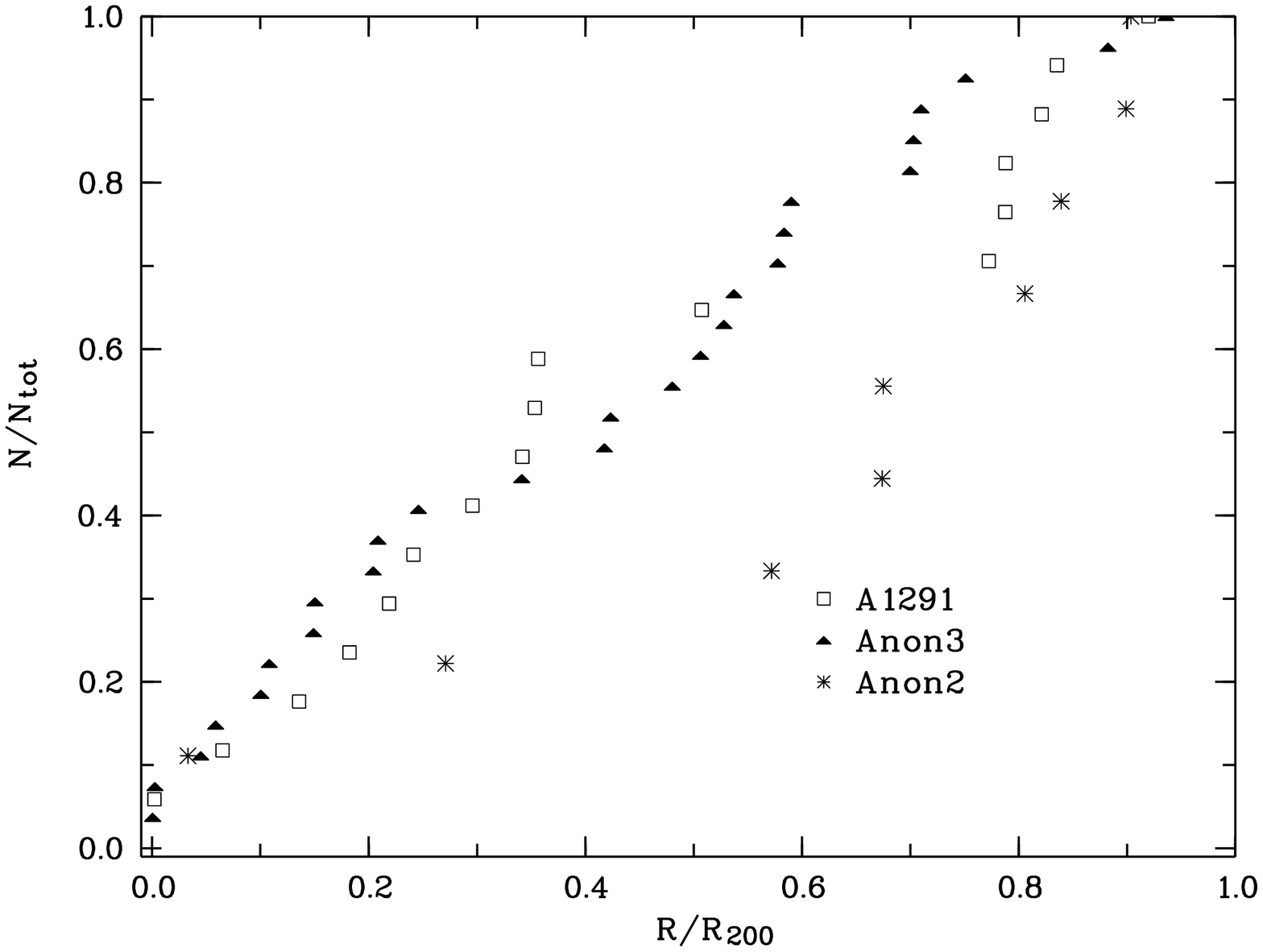,width=14cm,angle=-90,%
bbllx=69pt,bblly=133pt,bburx=520pt,bbury=730pt,clip=}}
\caption[]{(c)
}
\end{figure}

\begin{figure}[*]
\centerline{\psfig{figure=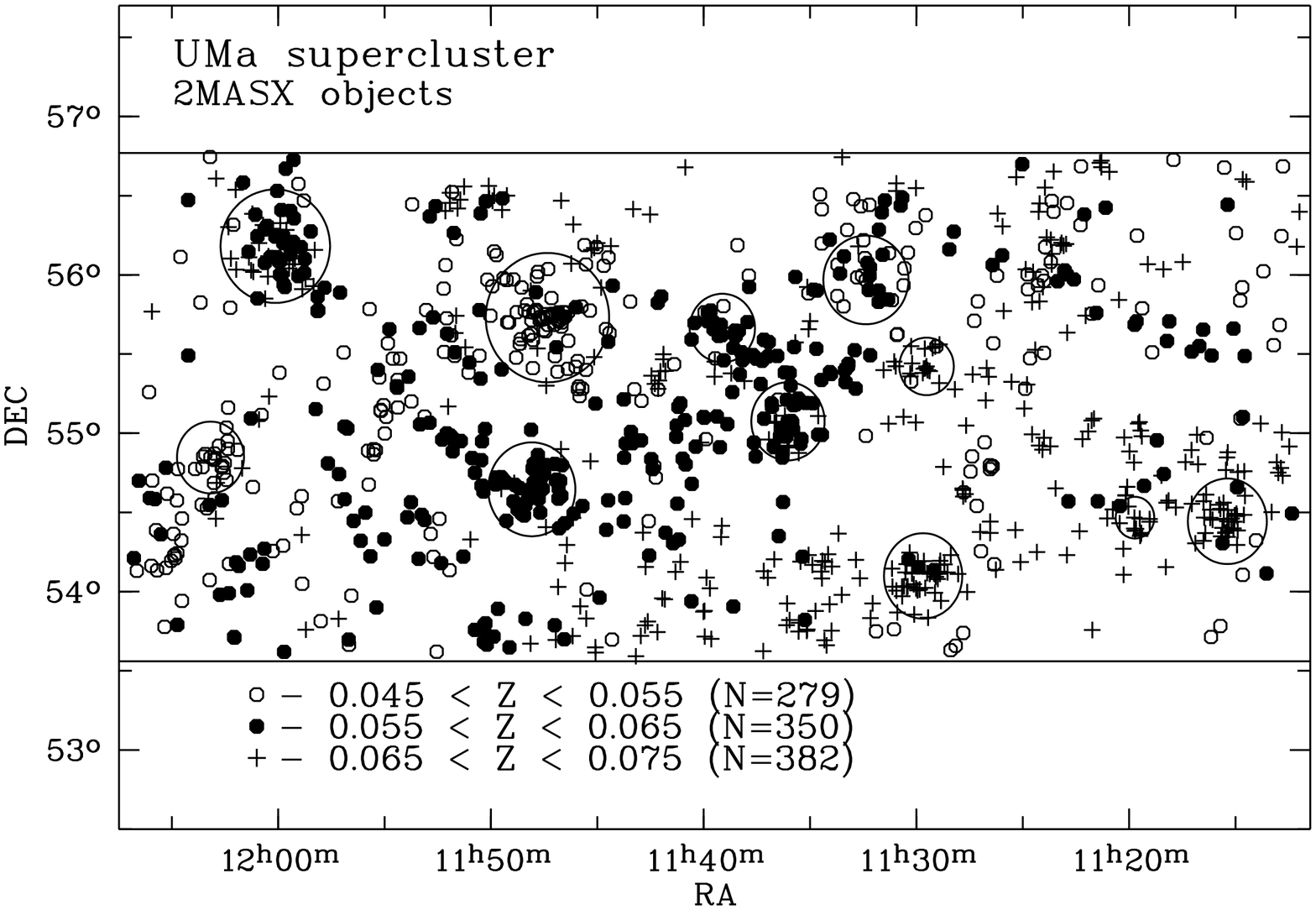,width=16cm,angle=-90,%
bbllx=52pt,bblly=50pt,bburx=552pt,bbury=761pt,clip=}}
\caption[]{Large-scale structure in the region of the Ursa Major
supercluster. The clusters are indicated by large circles with radii
equal to $R_{200}$.
}
\end{figure}

\begin{figure}[*]
\centerline{\psfig{figure=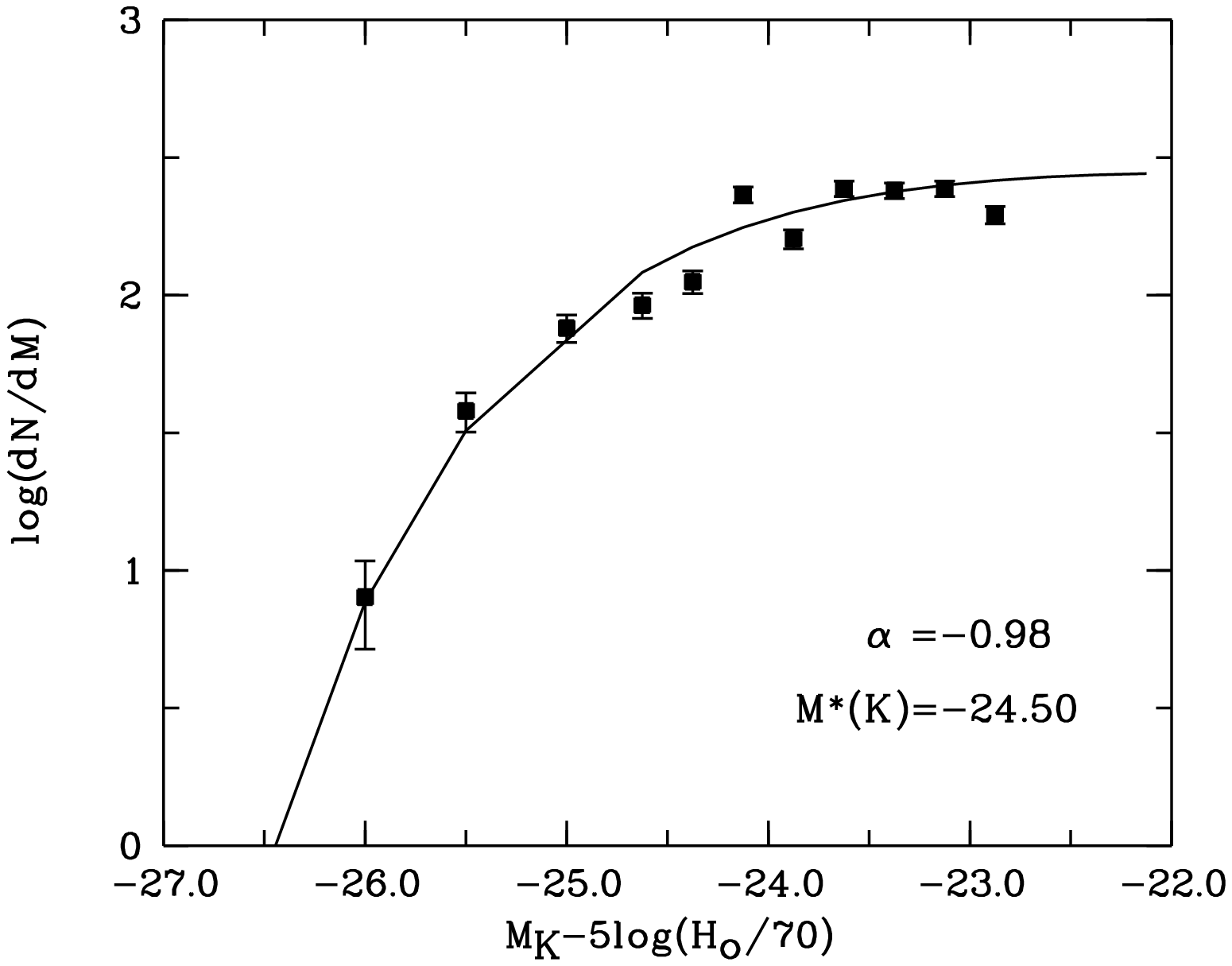,width=14cm,angle=-90,%
bbllx=112pt,bblly=131pt,bburx=488pt,bbury=598pt,clip=}}
\caption[]{Combined luminosity function for the UMa cluseters within
$R_{200}$. The errors are defined as $\sqrt{n}$, where $n$ is the
number of objects in the bin. The curve indicates the resulting
Schechter function (Schechter 1976), its  parameters ($M^{*}_{K}$
and $\alpha$) are given.
}
\end{figure}

\begin{figure}[*]
\centerline{\psfig{figure=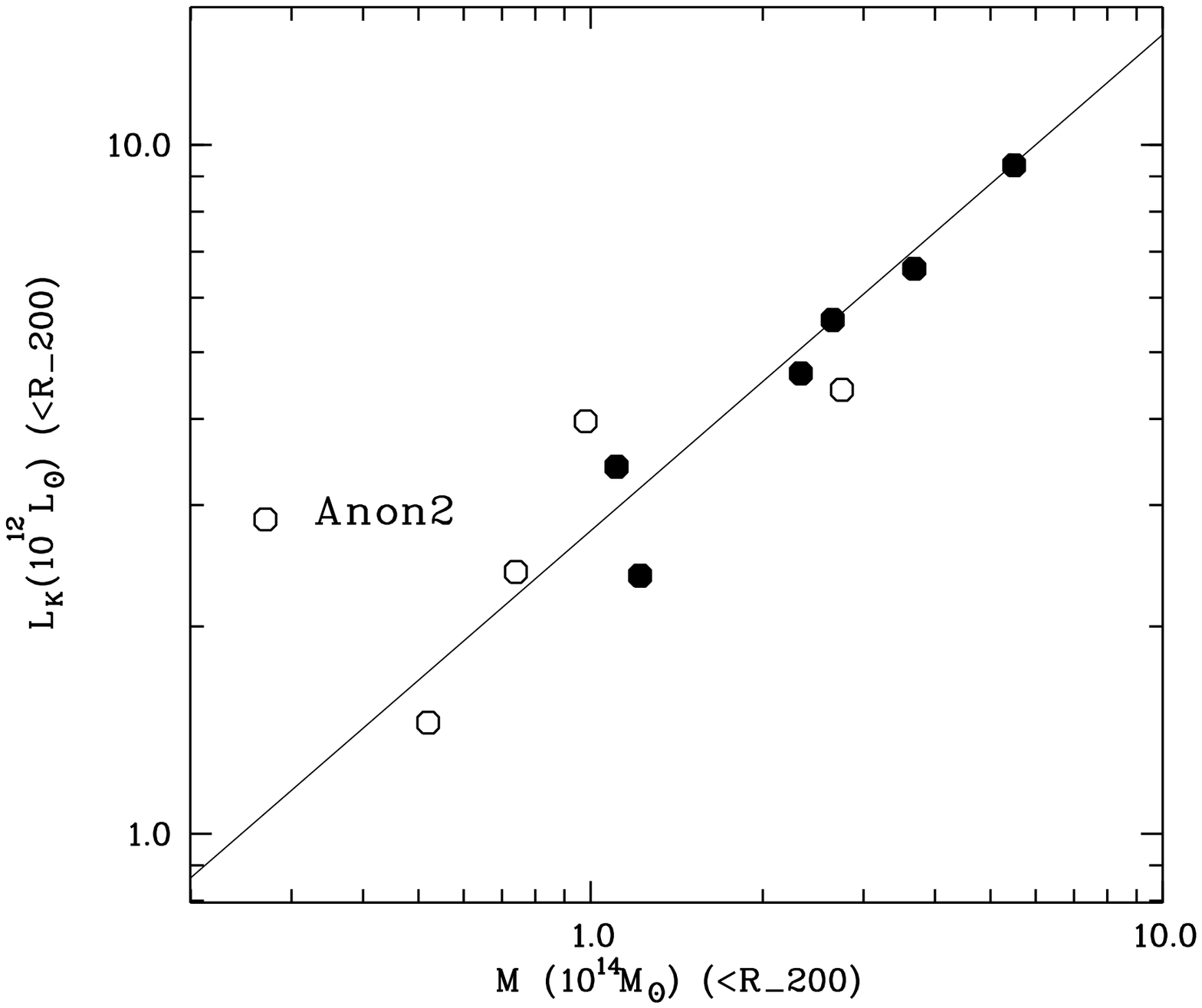,width=14cm,angle=-90,%
bbllx=98pt,bblly=62pt,bburx=535pt,bbury=591pt,clip=}}
\caption[]{
Relation between the mass within $R_{200}$ and the K-band luminosity
for the UMa clusters of galaxies: the filled circles represent the
clusters A1270, A1291, A1377, A1318, A1383, A1436; the open circles
represent Anon1, Anon2, Anon3, Anon4, Sh166; the line indicates the
relation derived by Lin et al. (2004) with a slope of 0.72.
}
\end{figure}

\begin{figure}[*]
\centerline{\psfig{figure=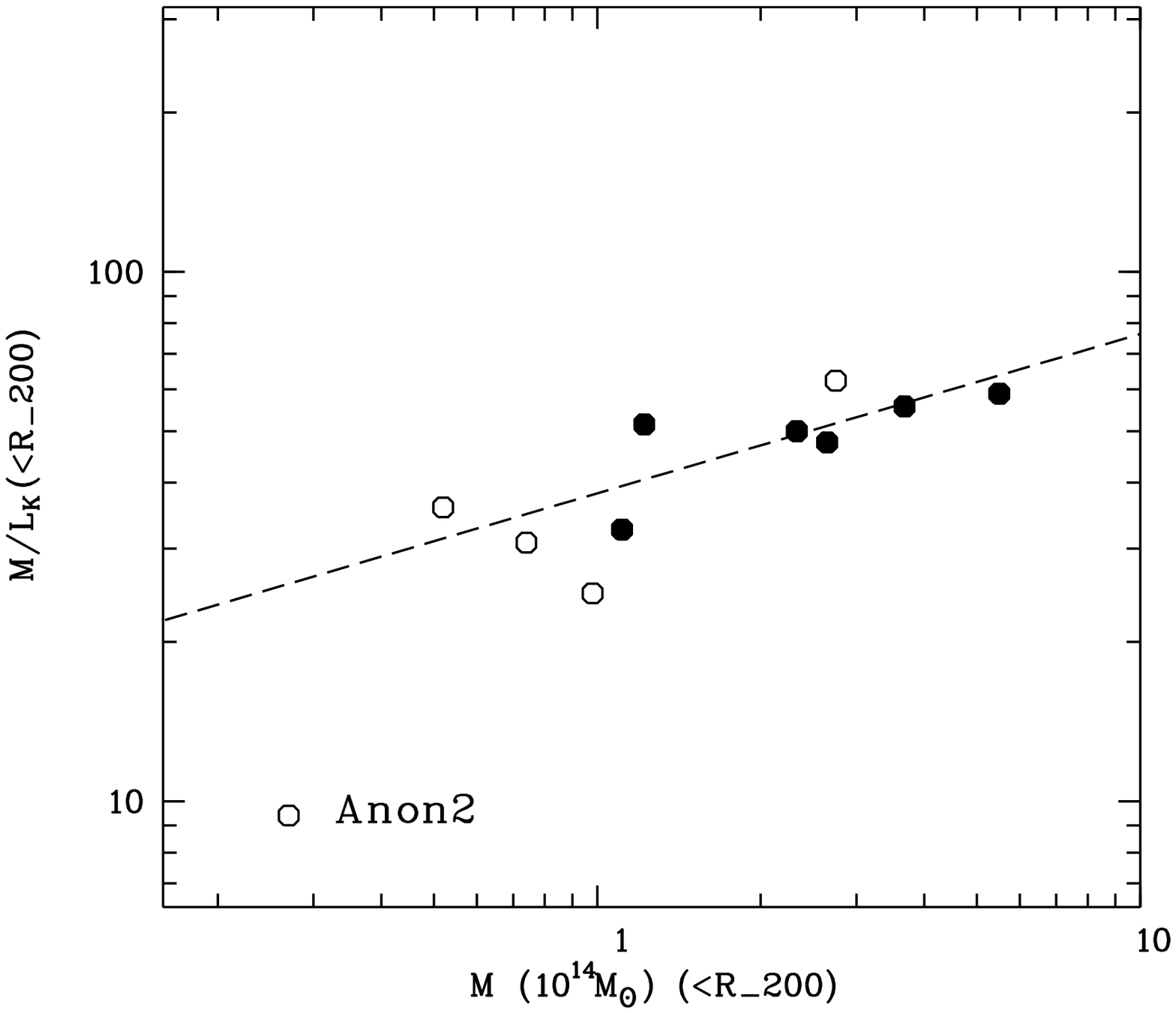,width=14cm,angle=-90,%
bbllx=100pt,bblly=144pt,bburx=555pt,bbury=649pt,clip=}}
\caption[]{Mass-to-luminosity ratio versus mass within $R_{200}$ for
the UMa clusters of galaxies. The dashed line indicates the relation
with a slope of 0.26 (Lin et al. 2004). The notation is the same as
that in Fig. 4.
}
\end{figure}
\end{document}